\documentclass[aps,prd,notitlepage,showpacs,nofootinbib,superscriptaddress]{revtex4-1}

\usepackage[utf8]{inputenc} 

\usepackage{amsmath}
\usepackage{amssymb}
\usepackage{float}
\usepackage{comment}
\usepackage{slashed}
\usepackage[normalem]{ulem}
\usepackage{bbold}
\usepackage{cancel}
\usepackage{graphicx}
\usepackage{multirow}
\usepackage{rotating}
\usepackage{caption}
\usepackage{subcaption}
\usepackage{tabulary}

\newcolumntype{K}[1]{>{\centering\arraybackslash}m{#1}}
\def\gsim{\raise0.3ex\hbox{$\;>$\kern-0.75em\raise-1.1ex\hbox{$\sim\;$}}}
\def\lsim{\raise0.3ex\hbox{$\;<$\kern-0.75em\raise-1.1ex\hbox{$\sim\;$}}}

\usepackage[usenames,dvipsnames]{color}

\newcommand {\ignore}[1]{}

\usepackage[colorinlistoftodos]{todonotes}
\usepackage[colorlinks=true,linkcolor=darkred,citecolor=darkred,urlcolor=darkred, pdfborder={0 0 0}]{hyperref}
\usepackage[normalem]{ulem}

\usepackage{caption}
\usepackage{subcaption}
\definecolor{linkcolor}{rgb}{0,0,0.5}

\definecolor{darkgreen}{rgb}{0,0.5,0}
\definecolor{darkred}{rgb}{0.6,0,0}
\definecolor{brown}{rgb}{0.59, 0.29, 0.0}
\definecolor{mightnightblue}{RGB}{25,25,112}

\def\sm{Standard Model }
\def\vev#1{\left\langle #1\right\rangle}
\def\SM{$\mathrm{SU(3)_c \otimes SU(2)_L \otimes U(1)_Y}$ }
\def\EW{$\mathrm{SU(2)_L \otimes U(1)_Y}$ }

\def\znbb {$\rm 0\nu\beta\beta$ }

\def\MR{\mathbf{M}_R}
\def\Yl{\mathbf{Y}_\ell}
\def\Y{\mathbf{Y}}
\def\Ynu{\mathbf{Y}_\nu}
\def\Yf{\mathbf{Y}_f}

\def\Mnu{\mathbf{M}_\nu}
\def\Ml{\mathbf{M}_\ell}

\def\Ul{\mathbf{U}_\ell}
\def\U{\mathbf{U}}

\def\dmatm{\Delta m^2_{31}}
\def\dmsol{\Delta m^2_{21}}

\newcommand{\AddrAHEP}{%
  AHEP Group, Institut de F\'{i}sica Corpuscular --
  C.S.I.C./Universitat de Val\`{e}ncia, Parc Cient\'ific de Paterna.\\
 C/ Catedr\'atico Jos\'e Beltr\'an, 2 E-46980 Paterna (Valencia) - SPAIN}
\newcommand{\AddrCFTP}{%
	Departamento de F\'{\i}sica and CFTP, Instituto Superior T\'ecnico, Universidade de Lisboa, Av. Rovisco Pais 1, 1049-001 Lisboa, Portugal}
	\newcommand{\AddrIISERB}{Department of Physics,
 Indian Institute of Science Education and Research - Bhopal \\
 Bhopal Bypass Road, Bhauri, Bhopal, India}

\usepackage[force]{feynmp-auto}

\begin{document}


\title{\boldmath \color{BrickRed} Minimal scoto-seesaw mechanism with spontaneous CP violation}

\author{D.~M. Barreiros}\email{debora.barreiros@tecnico.ulisboa.pt}
\affiliation{\AddrCFTP}

\author{F.~R. Joaquim}\email{filipe.joaquim@tecnico.ulisboa.pt }
\affiliation{\AddrCFTP}

\author{R.~Srivastava}\email{rahul@iiserb.ac.in}
\affiliation{\AddrIISERB}

\author{J.~W.~F. Valle}\email{valle@ific.uv.es}
\affiliation{\AddrAHEP}

\begin{abstract}
\vspace{0.5cm}

We propose simple scoto-seesaw models to account for dark matter and neutrino masses with spontaneous CP violation. This is achieved with a single horizontal $\mathcal{Z}_8$ discrete symmetry, broken to a residual $\mathcal{Z}_2$ subgroup responsible for stabilizing dark matter. CP is broken spontaneously via the complex vacuum expectation value of a scalar singlet, inducing leptonic CP-violating effects. We find that the imposed $\mathcal{Z}_8$ symmetry pushes the values of the Dirac CP phase and the lightest neutrino mass to ranges already probed by ongoing experiments, 
so that normal-ordered neutrino masses can be cornered by cosmological observations and neutrinoless double beta decay experiments.
\end{abstract}

\maketitle
\noindent

\section{Introduction}
\label{Sect:intro}

The discovery that at least two neutrinos are massive, to comply with the results of neutrino oscillation experiments~\cite{McDonald:2016ixn,Kajita:2016cak}, opened up a new chapter in particle physics. By itself, however, this is not enough to single out a new ``Standard Model'' since there are still several other open questions to be answered. In particular, underpinning the symmetries~\cite{Ishimori:2010au} responsible for the observed pattern of neutrino oscillations remains a formidable task. Moreover, the basic understanding and interpretation of cosmological dark matter presents us with a comparable challenge~\cite{Bertone:2004pz}. The idea that dark matter mediates neutrino mass generation~\cite{Ma:2006km} has now become a paradigm for new physics, both within
~\cite{Ma:2008cu,Hirsch:2013ola,Merle:2016scw,Diaz:2016udz,Choubey:2017yyn,Bonilla:2018ynb,CentellesChulia:2019gic,Bonilla:2019hfb,CentellesChulia:2019xky,Restrepo:2019ilz,Avila:2019hhv}
as well as beyond the simplest \SM Standard Model (SM) gauge structure~\cite{Kang:2019sab,Leite:2019grf,CarcamoHernandez:2020ehn,Leite:2020bnb,Leite:2020wjl}. These ``Scotogenic models'' complement the seesaw mechanism by providing a central role to dark matter. A successful model should, in addition, address the origin of CP violation and provide some insight on the flavour structure seen in the oscillation data~\cite{deSalas:2020pgw}.

\begin{figure}[t!]
	\includegraphics[width=0.75\textwidth]{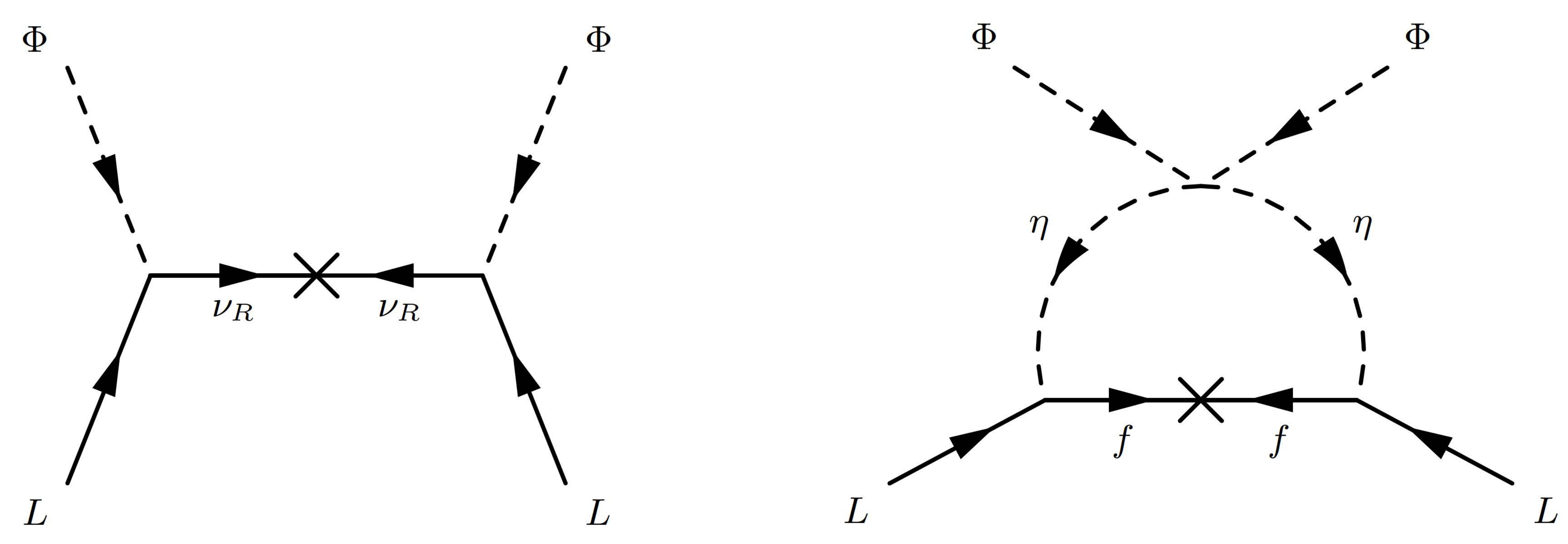}
	\caption{Neutrino mass generation in the minimal ``scoto-seesaw'' model. The left (right) diagram corresponds to the tree-level seesaw (one-loop scotogenic) contribution to the effective dimension-5 neutrino mass operator $LL\Phi\Phi$.}
	\label{minimaldiagrams}
\end{figure}

In Ref.~\cite{Rojas:2018wym}  the seesaw and scotogenic mechanisms have been cloned, as illustrated in fig.~\ref{minimaldiagrams}. In this way, a two-scale framework for neutrino oscillation was obtained, in which the atmospheric scale arises from the tree-level seesaw mechanism, while the solar scale has a radiative scotogenic origin. But there were no predictions for the flavour structure observed in the neutrino oscillation data. The latter can be addressed in the most straightforward way by using non-Abelian family symmetries.~\cite{Ma:2001dn,Babu:2002dz,Altarelli:2005yx,Ma:2014qra,deAnda:2019jxw,deAnda:2020pti,Chen:2020udk}. This would require extra symmetries beyond the scotogenic dark symmetry already inherent in the model. Instead, here we adopt a minimalistic approach to construct a scheme of neutrino mixing using \textsl{just one} symmetry. This single symmetry provides a ``texture zero''flavour structure for neutrinos, dark matter stability as well as spontaneous CP violation (SCPV). Moreover, we adopt a ``missing-partner'' framework with less ``right'' than ``left-handed'' neutrinos~\cite{Schechter:1979bn}, incorporated in a seesaw picture with Majorana neutrinos in~\cite{Schechter:1980gr}. The cases of one or two ``right-handed'' neutrinos predict the lightest neutrino to be massless, and hence a lower bound on the \znbb rate. These hold even if neutrinos are normal ordered, see~e.g.~\cite{Reig:2018ztc,Mandal:2019ndp,Leite:2019grf,Avila:2019hhv}.   

In this paper we extend the minimum ``scoto-seesaw'' dark matter model so as to address also the neutrino oscillation flavour structure and provide a spontaneous origin for leptonic CP violation.
Generalizing Ref.~\cite{Rojas:2018wym} we exploit the attractive features of combining the seesaw and scotogenic approaches within a scheme containing just two ``right-handed'' neutrinos. One way to do this is by using a simple an Abelian symmetry, provided at least two lepton families transform differently. This symmetry plays a role in stabilizing dark matter. To develop this approach in a realistic manner is messy, but has the virtue of being predictive and harboring spontaneous CP violation in the lepton sector.

This paper is organised as follows. In Section~\ref{sec:scot-dm-neutr} we start by presenting the minimal setup required to implement the aforementioned features of the model. In particular, we show how one can realise SCPV in the minimal scoto-seesaw context through complex vacuum expectation values (VEVs) of complex scalar singlets. We move on to construct the $\mathcal{Z}_8$ symmetric model, which is confronted with neutrino oscillation and neutrinoless double beta decay data in Sections~\ref{sec:neutr-mass-mixings} and \ref{sec:dbd}, respectively. Special attention is paid to the model predictions on the lightest neutrino mass $m_{\rm lightest}$ and Dirac CP-violation $\delta$. We also show that, in some cases, the model selects a specific $\theta_{23}$ octant. Finally, in Section~\ref{sec:concl} we draw our conclusions, and we provide details regarding the scalar sector of the model and SCPV in Appendix.

%
\section{Scotogenic dark matter, neutrino masses and spontaneous CP violation}
\label{sec:scot-dm-neutr}
%
Within the minimal scoto-seesaw model~\cite{Rojas:2018wym}, the atmospheric and solar neutrino mass scales are generated via the tree-level seesaw and the scotogenic mechanisms, respectively -- see fig.~\ref{minimaldiagrams}. The fermion sector contains one ``right-handed'' (RH) neutrino singlet $\nu_R$ and a fermion singlet $f$, while there is an extra scalar doublet $\eta$ besides the \sm Higgs doublet $\Phi$. Under the {\em dark} $\mathcal{Z}_2$ symmetry, $f$ and $\eta$ are odd, guaranteeing the stability of the dark matter (DM) particle, i.e. the lightest dark state. With such field content, the most general lepton Yukawa and mass terms read:
\begin{equation}
	-\mathcal{L}= \overline{L}\Yl\Phi e_R+\overline{L}\Ynu^*\tilde{\Phi}\nu_R+\dfrac{1}{2}M_R\overline{\nu_R}\nu_R^c+\overline{L}\Yf^*\tilde{\eta}f+\dfrac{1}{2}M_f\overline{f}f^c+{\rm H.c.}\,,
	\label{L1}
\end{equation}
where $L_i=(\nu_i\; \ell_i)^T$ and $e_{Ri}$ $(i = 1,2,3)$ are the left-handed (LH) and right-handed (RH) lepton doublets and singlets, respectively. The scalars $\Phi=(\phi^+\; \phi^0)^T$ and $\eta=(\eta^+\; \eta^0)^T$ are both $SU(2)_L$ doublets. In this notation, $\tilde{\Phi}=i\sigma_2\Phi^\ast$ and $\tilde{\eta}=i\sigma_2\eta^\ast$. $\Ynu$ and $\Yf$ are (complex) $3\times 1$ Yukawa coupling matrices, and $M_{R,f}$ are the $\nu_R$ and $f$ masses. The  resulting effective neutrino mass matrix reads~\cite{Rojas:2018wym},
\begin{align}
\Mnu=-v^2\dfrac{\Ynu^{}\Ynu^T}{M_R}+\mathcal{F}(M_f,m_{\eta_\text{R}},m_{\eta_\text{I}})M_f\Yf\Yf^T,
\label{Mnuscotoseesaw_min}
\end{align}
where the first term is the (tree-level) seesaw contribution, while the second accounts for the scotogenic radiative corrections (left and right diagrams in fig.~\ref{minimaldiagrams}, respectively),
where 
\begin{align}
    \mathcal{F}(M_f,m_{\eta_\text{R}},m_{\eta_\text{I}})=\dfrac{1}{32\pi^2}\left[\dfrac{m_{\eta_\text{R}}^2\log\left(M_f^2/m_{\eta_\text{R}}^2\right)}{M_f^2-m_{\eta_\text{R}}^2}-\dfrac{m_{\eta_\text{I}}^2\log\left(M_f^2/m_{\eta_\text{I}}^2\right)}{M_f^2-m_{\eta_\text{I}}^2}\right]\,.
\end{align}
Here, $m_{\eta_R}$ and $m_{\eta_I}$ are the masses of the real and imaginary parts of $\eta=\eta_{\rm R}+i \eta_{\rm I}$. Clearly this model can accommodate neutrino data, predicting the lightest neutrino to be massless. Notice however that, beyond this, there are no other predictions that can restrict, say, the oscillation parameters.

\begin{figure}[t!]
	\includegraphics[width=0.75\textwidth]{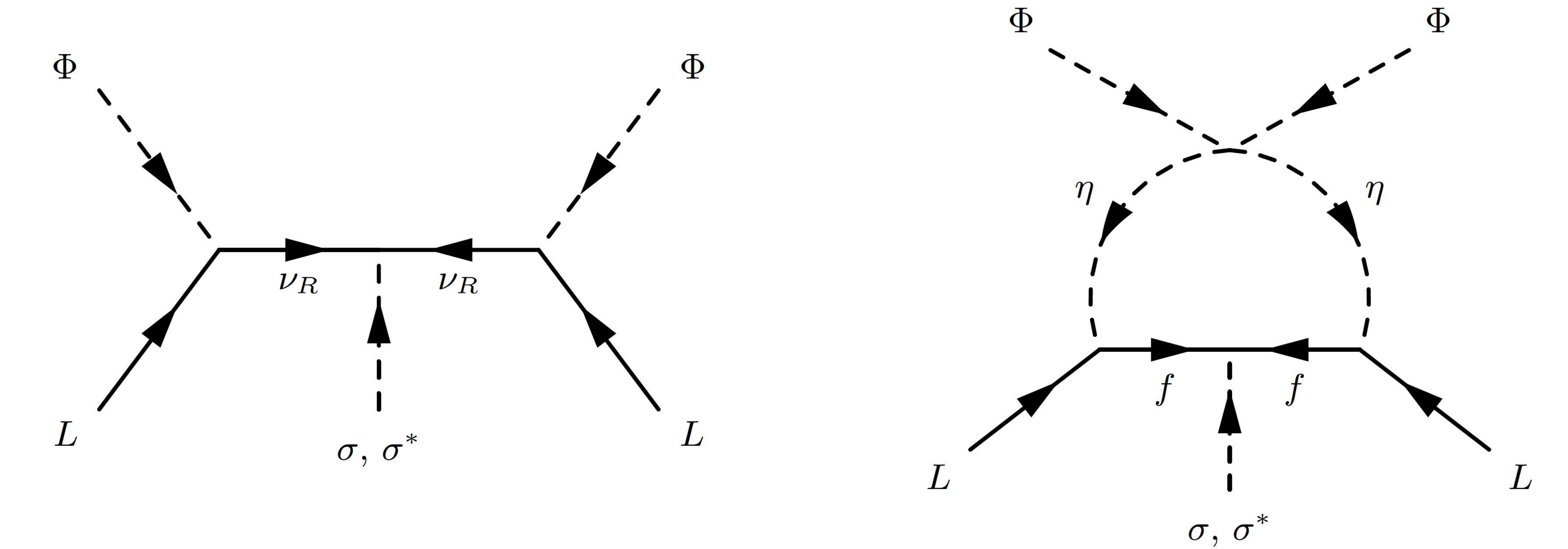}
	\caption{Neutrino mass diagrams in the minimal scotogenic seesaw model with LCPV induced by the couplings with a complex scalar singlet $\sigma$ and its conjugate $\sigma^\ast$.}
	\label{minimaldiagrams2}
\end{figure}

An interesting scenario may be realized by requiring that \eqref{L1} is invariant under CP and, thus, all couplings and masses are real. Clearly, in this case there is no leptonic CP violation (LCPV). However, introducing a complex scalar singlet $\sigma$ which acquires a complex VEV $\vev{ \sigma} = u e^{i\theta}$ will potentially induce LCPV
via couplings of $\nu_R$ and $f$ with $\sigma$ and $\sigma^\ast$~\cite{Branco:1999fs,Branco:2011zb}. In the most general case, these couplings are %
\begin{align}
\dfrac{1}{2}(y_R\sigma+\tilde{y}_R \sigma^\ast)\overline{\nu_R}\nu_R^c+\dfrac{1}{2}(y_f\sigma+\tilde{y}_f \sigma^\ast)\overline{f}f^c+{\rm H.c.}\,.
\label{L2}
\end{align}
Taking now into account that $\langle \sigma \rangle = u e^{i\theta}$, and expressing $M_{R,f}$ in \eqref{L1} as $M_{R,f}=|M_{R,f}|e^{i\theta_{R,f}}$, one has for the effective neutrino mass matrix \eqref{Mnuscotoseesaw_min}:
\begin{align}
\Mnu=-v^2 e^{i(\theta_f-\theta_R)}\dfrac{\Ynu^{}\Ynu^T}{|M_R|}+\mathcal{F}(|M_f|,m_{\eta_\text{R}},m_{\eta_\text{I}})|M_f|\Yf\Yf^T\,.
\label{Mnuscotoseesaw_min2}
\end{align}
with
\begin{align}
|M_{R,f}|^2=[\,y_{R,f}^2+\tilde{y}_{R,f}^2+2\,y_{R,f}\tilde{y}_{R,f}\cos(2\theta_{R,f})]\,u^2\;, \;
\tan{(\theta_f-\theta_R)}=\frac{(y_{f}\tilde{y}_{R}-y_{R}\tilde{y}_{f})\sin(2\theta)}{y_{R}y_{f}+\tilde{y}_{R}\tilde{y}_{f}
+(y_R\tilde{y}_{f}+y_f\tilde{y}_{R})\cos(2\theta)}\,,
\label{L2}
\end{align}
which shows that, in general, $\theta_f-\theta_R\neq 0$ and CP violation will be successfully communicated to the neutrino sector provided that $\theta \neq k\pi$ ($k=1,2,...$) and $y_{R,f}\neq \tilde{y}_{R,f}$.\footnote{Notice that, even if $\theta \neq k\pi$, one can have $\theta_f-\theta_R= 0$ if, for instance, both $\nu_R$ and $f$ acquire their mass just from coupling with $\sigma$ or $\sigma^\ast$.} The diagrammatic realization of what has just been described is depicted in fig.~\ref{minimaldiagrams2} where it can be clearly seen that CP-violation in $\Mnu$ requires a relative phase among the coefficients of the dimension-6 operators $LL\Phi \Phi \sigma$ and $LL\Phi \Phi \sigma^\ast$ stemming from the tree-level and scotogenic diagrams. Notice that, in this case, there are still nine parameters in \eqref{L1}, namely six real couplings in $\Ynu$ and $\Y_f$, the two masses $M_{R,f}$ and the relative phase $\theta_R-\theta_f$, which is higher than the seven effective neutrino parameters.

The minimal scoto-seesaw model, with just one copy of $\nu_R$ and $f$ fermions, does not offer a promising setup to implement neutrino predictions. In fact, any attempt to use an Abelian symmetry as a flavour symmetry to reduce the number of free parameters is likely to fail, as it will lead to incompatibility with neutrino oscillation data due to an unwanted extra massless neutrino or a vanishing leptonic mixing angle. Here, we extend the minimal scoto-seesaw model so as to implement a horizontal (flavour) symmetry which breaks to the {\em dark} $\mathcal{Z}_2$ subgroup responsible for dark matter stability. We do this by adding another $\nu_R$ in the fermion sector. Our template is then the (3,2) seesaw scheme~\cite{Schechter:1980gr}. In this case we add just two heavy singlet neutral leptons $\nu_{1R}$ and $\nu_{2R}$ to the Standard Model, along with the minimal dark sector with $\eta$ and $f$.

The lepton Yukawa and mass Lagrangian is identical to \eqref{L1} with $\nu_R$ replaced by $(\nu_{1R}\,\,\nu_{2R})^T$, and $M_R$ replaced by a $2\times 2$ symmetric matrix $\MR$. In this case, the most constraining patterns for the Yukawa and mass matrices compatible with neutrino data are~\cite{Barreiros:2018ndn,Barreiros:2018bju,Barreiros:2020mnr}:
%
%
\begin{align}
\Ynu=\begin{pmatrix}
\times&0\\
0&\times\\
\times&0\\
\end{pmatrix},\quad\Yf=\begin{pmatrix}
\times\\
0\\
\times\\
\end{pmatrix},
\quad \Yl=\begin{pmatrix}
\times&0&\times\\
0&\times&0\\
\times&0&\times\\
\end{pmatrix},\quad 
    \MR=\begin{pmatrix}
    0&\times\\
    \times&\times
    \end{pmatrix}\,
\label{viable_text_1}
\end{align}
with $\Ynu$ now promoted to a $3\times 2$ matrix. As will be seen later, the texture zeros in the matrices of \eqref{viable_text_1} can be moved up-and-down, as long as the total number of texture zeros is kept constant. At this point, we seek for the simplest symmetry which leads to the above patterns and allows for CP to be broken spontaneously through the VEV of the complex scalar singlet $\sigma$, as discussed above. On the other hand, since $\sigma$ couples to $\nu_{1R}$, $\nu_{2R}$ and $f$, the complex phase $\theta$ in $\vev{ \sigma} = u e^{i \theta}$ is communicated to the effective light neutrino mass matrix. This gives rise to CPV in the neutrino sector~\footnote{Notice that, since $\eta$ will not acquire a VEV, spontaneous CPV cannot be achieved via the \sm Higgs doublet VEVs.}. The relevant couplings which contribute to $\MR$ and $M_f$ in eq.~\eqref{L1} are thus
\begin{equation}
\frac{1}{2}\left({\rm \bf Y}_R \sigma+\widetilde{\Y}_R \sigma^*\right)\overline{\nu_R}\nu_R^c+\frac{1}{2}(y_f\sigma+\tilde{y}_f\sigma^*)\overline{f}f^c+{\rm H.c.}\,,
\label{YRYf}
\end{equation}
where $\Y_R$, $\widetilde{\Y}_R$ are now $2\times 2$ Yukawa matrices, while $y_f$ and $\widetilde{y}_f$ are complex numbers. In searching for an Abelian $\mathcal{Z}_N$ symmetry which realizes \eqref{viable_text_1} we must take into account that~\footnote{Continuous $U(1)$ symmetries will not be suitable for our purposes, since the presence of the $\sigma^4$ term requires $\sigma$ to have no $U(1)$ charge. As a result $M_R$ would receive both bare and $\sigma$-induced contributions.}
\begin{itemize}

\item The $\Ynu$ pattern \eqref{viable_text_1} requires the charges of $\nu_R^1$ and $\nu_R^2$ to be different and the mixing between the $\nu_R$ and $f$ to be forbidden. So we take a symmetry with at least three distinct non-trivial charges, i.e. $\mathcal{Z}_N$, with $N\geq4$.
    
\item In order to guarantee that $\MR$ has a texture zero, one of the $\nu_R^{1,2}$ must have a non-trivial (even) charge under the $\mathcal{Z}_N$ symmetry ($N\geq4$). In $\mathcal{Z}_4$, that does not happen since the only two even charges are 0 and 2, which multiplied by two gives zero or $N$ and hence will not lead to texture zeros. 
    
\item The dark sector particles (in the scotogenic loop) should not mix with the remaining particles. This can be achieved if their $\mathcal{Z}_N$ charges are distinct from those of the other particles. In addition, to ensure the stability of the lightest dark particle we require that a) the scalar $\sigma$ carries a non-trivial $\mathcal{Z}_N$ charge, so that its VEV breaks $\mathcal{Z}_N \to \mathcal{Z}_2$, and b) under the residual $\mathcal{Z}_2$ subgroup all dark sector particles have ``odd'' charge, while other particles are ``even''. All scalar fields with odd charges must not acquire VEV, so as to preserve the residual dark symmetry. Given these conditions one can see that ``odd'' $\mathcal{Z}_N$ groups like $\mathcal{Z}_5$ will not work. This is due to the fact that for odd $\mathcal{Z}_N$ groups if we assign an odd charge to a given dark field then its Hermitian conjugate will always have an even charge under the odd $\mathcal{Z}_N$, hence it can mix with other particles with even $\mathcal{Z}_N$ charge. In conclusion, for odd $\mathcal{Z}_N$ groups, one cannot ensure dark matter stability~\cite{Bonilla:2018ynb,Bonilla:2019hfb}.
       
\item The symmetry must allow for quartic $\sigma^4$ terms in the scalar potential, though it may be softly broken by $\sigma^2$ terms. The $\mathcal{Z}_5$, $\mathcal{Z}_6$ and $\mathcal{Z}_7$ groups do not work since invariance of the $\sigma^4$, needed to implement SCPV, cannot be ensured.
\end{itemize}

\begin{figure}[t!]
	\includegraphics[width=0.75\textwidth]{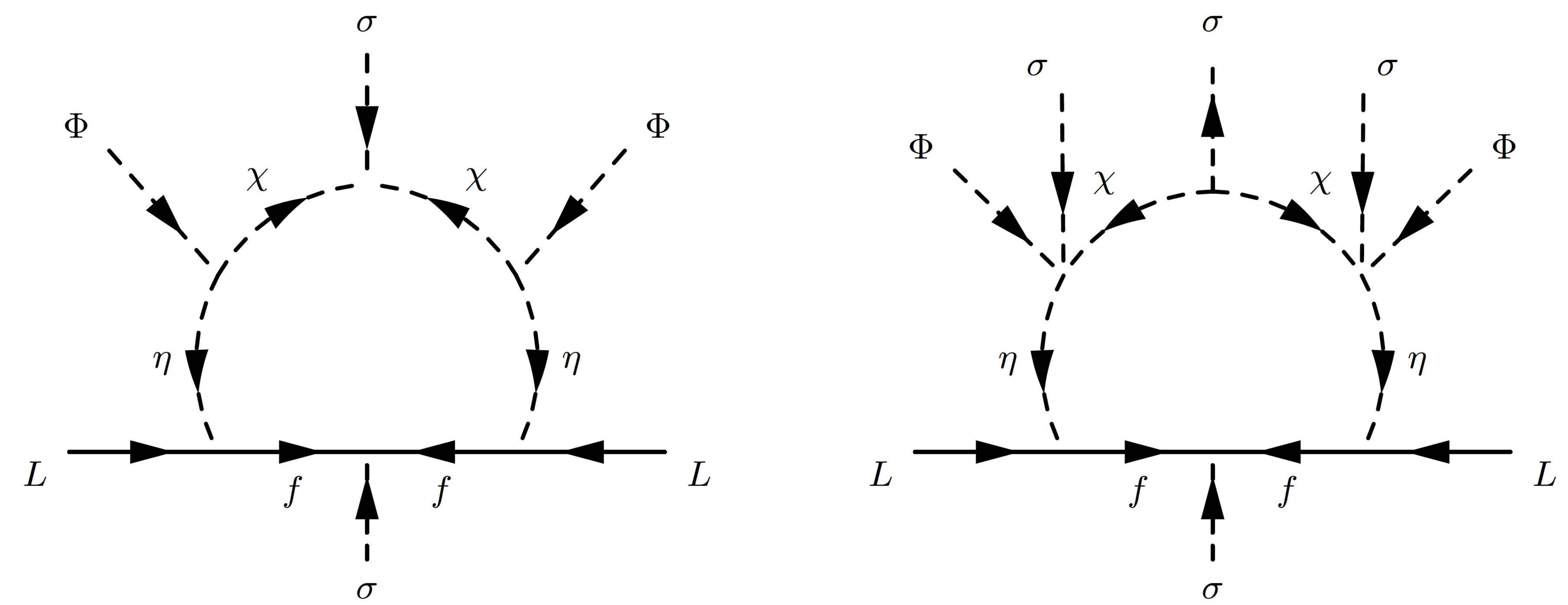}
	\caption{One-loop diagrams that contribute radiatively to neutrino mass.}
	\label{scotoloopsZ8}
\end{figure}

Putting things together we conclude that the minimal required symmetry to reproduce the textures in \eqref{viable_text_1} is $\mathcal{Z}_8$. In table~\ref{modelZ8}, we show the matter content and possible charge assignments for our models. As shown in that table, there are in fact three possible assignments which we label as $\mathcal{Z}^{e-\mu}_8$, $\mathcal{Z}^{\mu-\tau}_8$ and $\mathcal{Z}^{e-\tau}_8$. Hence one can consider three distinct models. Notice that just with the scalars $\Phi$, $\eta$ and $\sigma$, only the tree-level seesaw contributes to the effective neutrino mass matrix. This follows from the fact that the scotogenic loop of fig.~\ref{minimaldiagrams} requires a $\mathcal{Z}_8$ explicit breaking term $(\phi^\dagger \eta)^2$. Moreover, as discussed below, just with the tree-level seesaw contribution, the complex phase of the $\sigma$ VEV does not lead to leptonic CP violation. Hence, we add a second dark sector complex singlet $\chi$ transforming under $\mathcal{Z}_8$ as shown in table~\ref{modelZ8}. The scalar potential of the model is then
\begin{align}
	V=\;&m_\Phi^2\Phi^\dagger \Phi+m_\eta^2\eta^\dagger\eta+m_\sigma^2\sigma^\ast \sigma+m_\chi^2\chi^\ast \chi+\dfrac{\lambda_1}{2}(\Phi^\dagger \Phi)^2+\dfrac{\lambda_2}{2}(\eta^\dagger \eta)^2+\dfrac{\lambda_{3}}{2}(\sigma^\ast \sigma)^2+\dfrac{\lambda_4}{2}(\chi^\ast \chi)^2+\lambda_{5}(\Phi^\dagger \Phi)(\eta^\dagger \eta)+\nonumber\\
	&+\lambda'_{5}(\Phi^\dagger \eta)(\eta^\dagger \Phi)+\lambda_{6}(\Phi^\dagger\Phi)(\sigma^\ast\sigma)+\lambda_{7}(\Phi^\dagger\Phi)(\chi^\ast\chi)+\lambda_{8}(\eta^\dagger\eta)(\sigma^\ast\sigma)+\lambda_{9}(\eta^\dagger\eta)(\chi^\ast\chi)+\lambda_{10}(\sigma^\ast\sigma)(\chi^\ast\chi)+\nonumber\\
	&+\left(\dfrac{\lambda'_{3}}{4}\sigma^4+\dfrac{m'^2_\sigma}{2}\sigma^2+\mu_1\chi^2\sigma+\mu_2\eta^\dagger \Phi \chi^\ast+\lambda_{11}\eta^\dagger\Phi\sigma\chi+\rm{H.c.}\right).
\end{align}
where all parameters are real so as to ensure CP invariance at the Lagrangian level. Notice that the three last terms allow for the one-loop scotogenic diagrams in fig.~\ref{scotoloopsZ8}. These contribute to neutrino mass generation at lowest dimension of the neutrino mass operators.

We now turn to the issue of SCPV. As shown in the Appendix, the minimisation of the scalar potential can lead to a VEV configuration with
$\vev{ \eta^0}=0$, $\vev{ \chi}=0$, $\vev{ \phi^0}=v$ and $\vev{ \sigma}=u e^{i\theta}$, provided:
\begin{align}
  \label{eq:scpv}
m_\Phi^2=-\dfrac{\lambda_1}{2}v^2-\dfrac{\lambda_6}{2}u^2\;, \quad m_\sigma^2=-\dfrac{\lambda_6}{2}v^2-\dfrac{\lambda_3-\lambda_3'}{2}u^2\;,\quad\cos (2\theta)=-\dfrac{m'^2_\sigma}{u^2\lambda'_3}\,.
\end{align}
Note that the presence of the scalar singlet $\chi$ plays a key role in our model to prevent the explicit breaking of $\mathcal{Z}_8$ in the scotogenic loop.
\begin{table}[t!]
\setlength{\tabcolsep}{-1pt}
	\centering
	\begin{tabular}{| K{0.5cm} || K{1.5cm} | K{3cm} |  K{4cm} | K{4cm} | K{4cm} |}
		\hline 
&Fields&\EW&$\mathcal{Z}^{e-\mu}_8 \to \mathcal{Z}^D_2$ & $\mathcal{Z}^{\mu-\tau}_8 \to \mathcal{Z}^D_2$ & $\mathcal{Z}^{e-\tau}_8 \to \mathcal{Z}^D_2$ \\
		\hline \hline
		\multirow{6.5}{*}{\begin{turn}{90} Fermions \end{turn}} 
&$L_e,e_R$&($\mathbf{2}, {-1/2}$),\,($\mathbf{1}, 0$)&{$\omega^6\equiv -i$} $\to$ $+1$	
& {$\omega^0 \equiv 1$}  $\to$  $+1$ 
& {$\omega^6\equiv -i$} $\to$ $+1$ \\
&$L_{\mu},\mu_R$&($\mathbf{2}, {-1/2}$),\,($\mathbf{1}, 0$)&{$\omega^6\equiv -i$} $\to$ $+1$
& {$\omega^6\equiv -i$}  $\to$ $+1$     
 & {$\omega^0 \equiv 1$} $\to$  $+1$\\
& $L_\tau,\tau_R$&($\mathbf{2}, {-1/2}$),\,($\mathbf{1}, 0$)&  {$\omega^0 \equiv 1$}$\to$  $+1$ 
&  {$\omega^6\equiv -i$}  $\to$  $+1$    
& {$\omega^6\equiv -i$} $\to$ $+1$ \\
&$\nu_R^{1}$&($\mathbf{1}, {0}$)&{$\omega^6\equiv -i$} $\to$ $+1$ 
&  {$\omega^6\equiv -i$}  $\to$  $+1$   
& {$\omega^6\equiv -i$} $\to$  $+1$   \\
&$\nu_R^{2}$&($\mathbf{1}, {0}$)&{$\omega^0 \equiv 1$} $\to$ $+1$ 
&  {$\omega^0 \equiv 1$} $\to$  $+1$ 
& {$\omega^0 \equiv 1$}  $\to$  $+1$ \\
&$f$&($\mathbf{1}, {0}$)& {$\omega^3$}   $\to$  $-1$ 
& {$\omega^3$}   $\to$  $-1$
& {$\omega^3$}   $\to$ $-1$\\
		\hline \hline
		\multirow{4.5}{*}{\begin{turn}{90} Scalars \end{turn}}
&$\Phi$&($\mathbf{2}, {1/2}$)& {$\omega^0 \equiv 1$} $\to$ $+1$ 
& {$\omega^0 \equiv 1$} \, \, $\to$ \, \, $+1$
& {$\omega^0 \equiv 1$}$\to$  $+1$\\
&$\sigma$&($\mathbf{1}, {0}$)& {$\omega^2 \equiv i$}  $\to$  $+1$ 
&  {$\omega^2 \equiv i$} $\to$  $+1$ 
& {$\omega^2 \equiv i$}  $\to$ $+1$ \\
&$\eta$&($\mathbf{2}, {1/2}$)&{$\omega^5$}    $\to$ $-1$
&{$\omega^5$}    $\to$  $-1$
& {$\omega^5$}   $\to$  $-1$\\		
&$\chi$&($\mathbf{1}, {0}$)&{$\omega^3$}  $\to$  $-1$
&{$\omega^3$} $\to$ $-1$
& {$\omega^3$}  $\to$ $-1$\\
		\hline
	\end{tabular}
	\caption{Matter content and charge assignments of the model. 
          Here $\omega^a = e^{2i\pi a/8}$ is the $a$-th power of the eight root of unity that defines the $\mathcal{Z}_8$ symmetry.}
	\label{modelZ8} 
\end{table}
An effective neutrino mass matrix analogous to the one in eq.~\eqref{Mnuscotoseesaw_min} can now be written as,
\begin{align}
\Mnu=-v^2\Ynu^{}\MR^{-1}\Ynu^T+\mathcal{F}(M_f,m_{S_i})M_f\Yf\Yf^T,
\label{Mnuscotoseesaw}
\end{align}
where $\Ynu$ is the Dirac neutrino Yukawa coupling matrix and $\Yf$ is the matrix of Yukawa-type couplings of the leptons to the dark fields $f$ and $\eta$. $\MR$ is the right-handed neutrino mass matrix and $M_f$ is the $f$ mass. The first term in eq.~\eqref{Mnuscotoseesaw} is the seesaw contribution, while the radiative corrections are given by the second term. The latter are characterized by the loop function $\mathcal{F}(M_f,m_{S_i})$, depending on the mass eigenstates resulting from the mixing of the neutral components of $\eta$ and $\chi$ (see Appendix). While in general, $\Ynu$, $\Yf$, and $\MR$ are complex $3\times 2$, $3\times 1$ and $2\times 2$ matrices, we assume that CP is conserved at the Lagrangian level. In this case the only form of breaking is spontaneous, dictated by the complex VEV $\vev{\sigma}=u e^{i\theta}$ of the $\sigma$ field in which the phase is determined through eq.~(\ref{eq:scpv}).

For definiteness let us focus on the model defined by the $\mathcal{Z}^{e-\tau}_8$ charge assignment in table~\ref{modelZ8}. Taking into account eqs.~\eqref{YRYf} and \eqref{Mnuscotoseesaw}, we find the following form for the Yukawa and mass matrices
\begin{align}
\Ynu=\begin{pmatrix}
x_1&0\\
0&x_2\\
x_3&0\\
\end{pmatrix},\quad\MR=\begin{pmatrix}
0&M_{12} \,e^{-i \theta}\\
M_{12}\,e^{-i \theta} &M_{22}\\
\end{pmatrix},\quad\Yf=\begin{pmatrix}
y_1\\
0\\
y_2\\
\end{pmatrix},
\quad \Yl=\begin{pmatrix}
w_1&0&w_2\\
0&w_3&0\\
w_4&0&w_5\\
\end{pmatrix}\,,
\label{matrix_struct}
\end{align}
where $M_{12}={(\Y_R)}_{12}u$ and $M_{22}$ are invariant bare masses. Due to the initially imposed CP invariance, the parameters $x_i$, $y_i$, $w_i$ and $M_{1,2}$ are real. Notice also that the $f$ mass term is $M_f e^{-i \theta}$ with $M_f=y_f u$. The resulting $\mathcal{Z}_8$-invariant $\Yl$ texture implies that one of the charged-leptons is decoupled from the remaining two in the symmetry basis. Thus, the contribution to lepton mixing coming from the charged lepton sector is non-trivial, being parametrized by a single angle $\theta_\ell$. Consequently, the unitary transformation which brings the $\ell_i$ fields to the mass-eigenstate basis is
\begin{align}
 \Ul= \begin{pmatrix}
    \cos\theta_\ell&0&\sin\theta_\ell\\
    0&1&0\\
    -\sin\theta_\ell&0&\cos\theta_\ell
    \end{pmatrix} \mathbf{P}_{ij}, 
    \label{Ul}
\end{align}
where $\theta_\ell$ is the mixing angle. The permutation matrices obey $ \mathbf{P}_{ij}= \mathbf{P}_{12},$ $\mathbb{1}$  or $ \mathbf{P}_{23}$ depending on whether the decoupled charged lepton is the electron ($\mathcal{Z}^{\mu-\tau}_8$ symmetry), muon ($\mathcal{Z}^{e-\tau}_8$ symmetry) or tau ($\mathcal{Z}^{e-\mu}_8$ symmetry), respectively, i.e.
\begin{align}
    \mathbf{P}_{12}=\begin{pmatrix}
    0&1&0\\
    1&0&0\\
    0&0&1
    \end{pmatrix},\;    
    \mathbf{P}_{23}=\begin{pmatrix}
    1&0&0\\
    0&0&1\\
    0&1&0
    \end{pmatrix}\,,
    \label{permutationmatrices}
\end{align}
and $\mathbb{1}$ the $3\times 3$ identity matrix. The charged-lepton and effective neutrino mass matrices in the charged-lepton physical basis are then
\begin{align}
\Ml'=\text{diag}(m_e,m_\mu,m_\tau) \;,\;\mathbf{M}_\nu'=\Ul^T \Mnu \Ul\,,
   \label{Mnu_clbasis}
\end{align}
with $\Mnu$ and $\U_\ell$ given in eqs.~\eqref{Mnuscotoseesaw} and \eqref{Ul}, respectively. Taking into account eqs.~\eqref{Mnuscotoseesaw} and \eqref{matrix_struct}, the effective neutrino mass matrix in the original symmetry basis is given by
\begin{align}
\Mnu =\left(
\begin{array}{ccc}
 \mathcal{F}(M_f,m_{S_i})\, M_f\, y_1^2+\dfrac{ v^2 M_{22}}{M_{12}^2} x_1^2e^{i \theta } \hspace*{0.5cm}& -\dfrac{v^2}{M_{12}}x_1 x_2 \hspace*{0.5cm} & \mathcal{F}(M_f,m_{S_i})\, M_f\, y_1 y_2+\dfrac{v^2 M_{22} }{M_{12}^2}x_1 x_3e^{i \theta } \\[0.5cm]
 \cdot& 0 \hspace*{0.5cm}& -\dfrac{v^2}{M_{12}} x_2 x_3 \\[0.5cm]
 \cdot& \cdot\hspace*{0.5cm}& \mathcal{F}(M_f,m_{S_i})\, M_f\, y_2^2+\dfrac{v^2 M_{22} }{M_{12}^2}x_3^2e^{i \theta } \\
\end{array}
\right),
\label{Mnuconst}
\end{align}
where the texture zero $(\Mnu)_{22}=0$ results from the assumed $\mathcal{Z}^{e-\tau}_8$ symmetry. It can be shown that the contribution of the scotogenic loop is crucial to ensure the existence of CP violation since, in its absence, the vacuum phase $\theta$ can be rephased away. In fact, by computing $\mathcal{J}_\text{CP}=\text{Tr}[(\Mnu\Mnu^\dagger)^*,\Ml\Ml^\dagger]$, one concludes that $\mathcal{J}_\text{CP}=0$ if $y_1=y_2=0$, i.e., if the loop contribution vanishes. Notice also that, since the charged-lepton sector has an unmixed state, the vanishing-element condition in $\Mnu$ remains after performing the $\U_\ell$ rotation of eq.~\eqref{Mnu_clbasis}. Namely, 
\begin{align}
(\mathbf{M}_\nu')_{ii}=0\,\,\text{for decoupled}~\,e_i\,,\label{c11}
\end{align}
with $i=1,2,3$ for $e,\,\mu$ and $\tau$, respectively.

\section{oscillation constraints on low-energy parameters}
\label{sec:neutr-mass-mixings}

In order to analyse the constraints imposed by \eqref{c11} on the low-energy neutrino parameters, we express $\mathbf{M}_\nu'$ in terms of the lepton mixing angles and neutrino masses. Namely,
\begin{align}
\mathbf{M}_\nu'=\U^*\mathbf{d}_m\U^\dagger\,,~~\mathbf{d}_m \equiv  \text{diag}(m_1,m_2,m_3)\,,
\label{Mnudiag}
\end{align}
where $m_i$ are the real and positive light neutrino masses and $\U$ is the lepton mixing matrix. Using the above equation one can reconstruct the effective neutrino mass matrix from the low-energy parameters.
\begin{table}[t!]
\centering
\setlength\extrarowheight{2pt}
\begin{tabular}{|l||K{3.5cm}|K{3.5cm}|}
\hline
Parameter&Best Fit $\pm1\sigma$&$3\sigma$ range\\
\hline\hline
$\theta_{12}\;(^{\circ})$ [NO] [IO]&$34.3\pm1.0$&$31.4-37.4$\\[0.15cm]
$\theta_{23}\;(^{\circ})$ [NO] &$48.79_{-1.25}^{+0.93}$&$41.63-51.32$\\
$\theta_{23}\;(^{\circ})$ [IO] &$48.79_{-1.30}^{+1.04}$&$41.88-51.30$\\[0.15cm]
$\theta_{13}\;(^{\circ})$ [NO] &$8.58_{-0.15}^{+0.11}$&$8.16-8.94$\\
$\theta_{13}\;(^{\circ})$ [IO] &$8.63_{-0.15}^{+0.11}$&$8.21-8.99$\\[0.15cm]
$\delta/\pi$ [NO] &$1.20_{-0.14}^{+0.23}$&$0.80-2.00$\\
$\delta/\pi$ [IO] &$1.54\pm 0.13$&$1.14-1.90$ \\[0.15cm]
$\Delta m_{21}^2\;(\times 10^{-5}\;\text{eV}^2)$  [NO] [IO]&$7.50_{-0.20}^{+0.22}$&$6.94-8.14$\\[0.15cm]
$|\Delta m_{31}^2|\;(\times 10^{-3}\;\text{eV}^2)$ [NO] &$2.56^{+0.03}_{-0.04}$&$2.46-2.65$\\
$|\Delta m_{31}^2|\;(\times 10^{-3}\;\text{eV}^2)$ [IO] &$2.46\pm0.03$&$2.37-2.55$\\
\hline
\end{tabular}
\caption{Neutrino oscillation parameters obtained from the global analysis of ref.~\cite{deSalas:2020pgw}.}
\label{datatable}
\end{table} 
After eliminating unphysical phases~\cite{Schechter:1980gr}, $\U$ can be parametrized in a symmetrical way~\cite{Rodejohann:2011vc} as
\begin{align}
\U=
\left( \begin{array}{c c c}
c_{12}c_{13}&s_{12}c_{13}e^{-i{\phi_{12}}}&s_{13}e^{-i{\phi_{13}}}\\
-s_{12}c_{23}e^{i{\phi_{12}}}-c_{12}s_{13}s_{23}e^{-i({\phi_{23}}-{\phi_{13}})}
&c_{12}c_{23}-s_{12}s_{13}s_{23}
e^{-i({\phi_{12}}+{\phi_{23}}-{\phi_{13}})}&c_{13}s_{23}e^{-i{\phi_{23}}}\\
s_{12}s_{23}e^{i({\phi_{12}}+{\phi_{23}})}-c_{12}s_{13}c_{23}e^{i{\phi_{13}}}
&-c_{12}s_{23}e^{i{\phi_{23}}}-
s_{12}s_{13}c_{23}e^{-i({\phi_{12}}-{\phi_{13}})}&c_{13}c_{23}\\
\end{array} \right)\,,
\label{Uparam}
\end{align}
where $\theta_{ij}$ ($i<j=1,2,3$) are the lepton mixing angles (with $s_{ij}\equiv\sin\theta_{ij}$, $c_{ij}\equiv\cos\theta_{ij}$). Here, $\delta=\phi_{13} -\phi_{12} - \phi_{23} $ is the Dirac CP-violating phase relevant for neutrino oscillations and $\phi_{12,13}$ are the two Majorana phases. For normal and inverted neutrino mass ordering (NO and IO, respectively) two neutrino masses are expressed in terms of the lightest neutrino mass $m_{\rm lightest}$ (which coincides with $m_1$ and $m_3$ for NO and IO, respectively), and the measured neutrino mass-squared differences $\dmsol=m_2^2-m_1^2$ and $\dmatm=m_3^2-m_1^2$ as
\begin{align}
   \text{NO:}&\quad m_2=\sqrt{m_{\rm lightest}^2+\dmsol},\quad m_3=\sqrt{m_{\rm lightest}^2+\dmatm}\quad,\\
   \text{IO:} &\quad m_1=\sqrt{m_{\rm lightest}^2+|\dmsol|},\quad m_2=\sqrt{m_{\rm lightest}^2+\dmsol+|\dmatm|}\quad\,.
\end{align}
The current experimental ranges extracted from global fits of neutrino oscillation data are shown in table~\ref{datatable}, from~\cite{deSalas:2020pgw}. 

\begin{figure}[t!]
\includegraphics[width=0.83\textwidth]{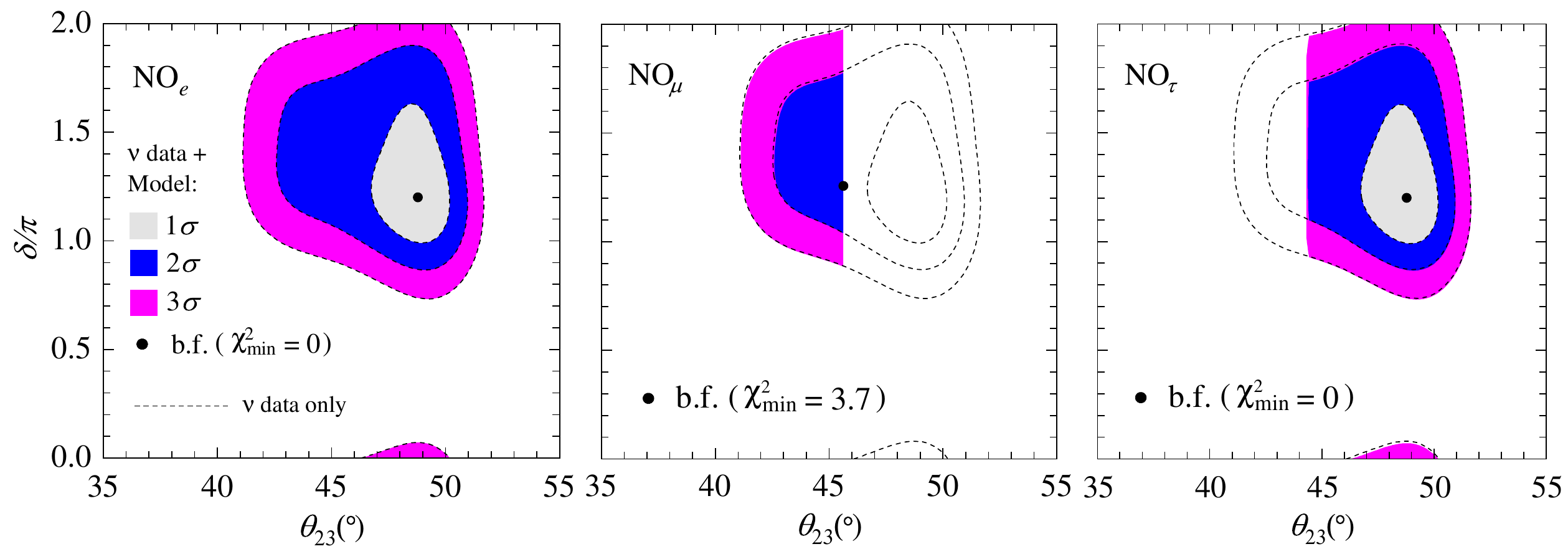}\\
\includegraphics[width=0.55\textwidth]{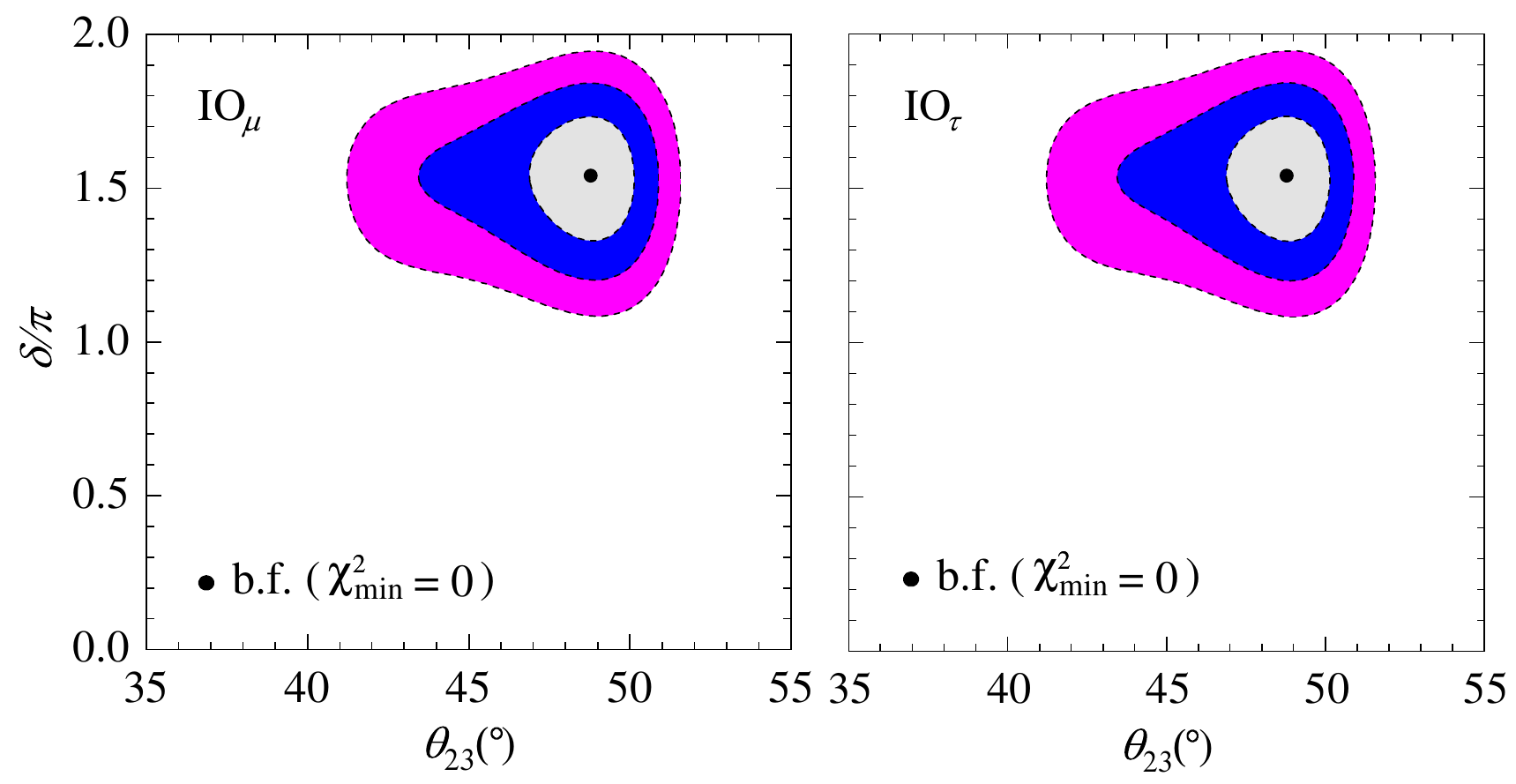}
\caption{Allowed regions in the plane ($\theta_{23}$,$\delta$) for NO$_{e,\mu,\tau}$ (upper plots) and IO$_{\mu,\tau}$ (lower plots) for the $\mathcal{Z}_8$ model considered. As discussed in the text the IO$_e$ case cannot be realized. The dashed lines delimit the allowed regions obtained considering only neutrino oscillation data, i.e. without input of the model.}
\label{results_t23}
\end{figure}

\begin{figure}[h!]
\includegraphics[width=0.8\textwidth]{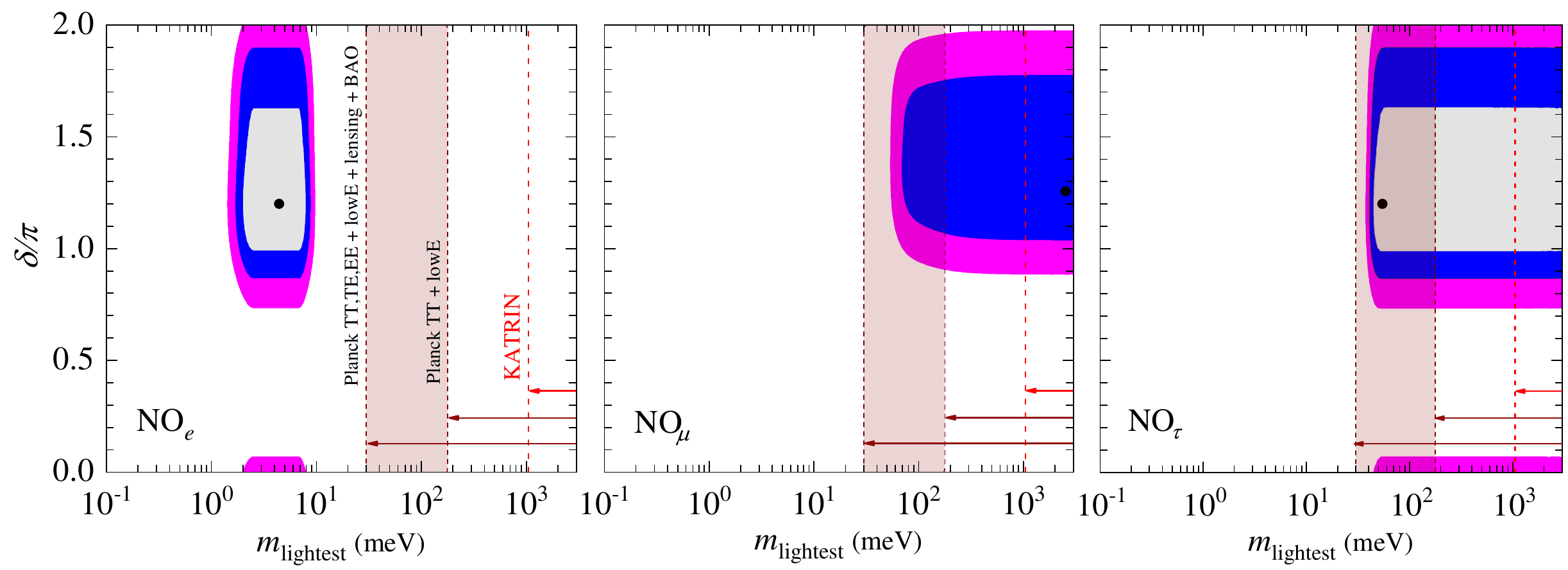}\\
\includegraphics[width=0.55\textwidth]{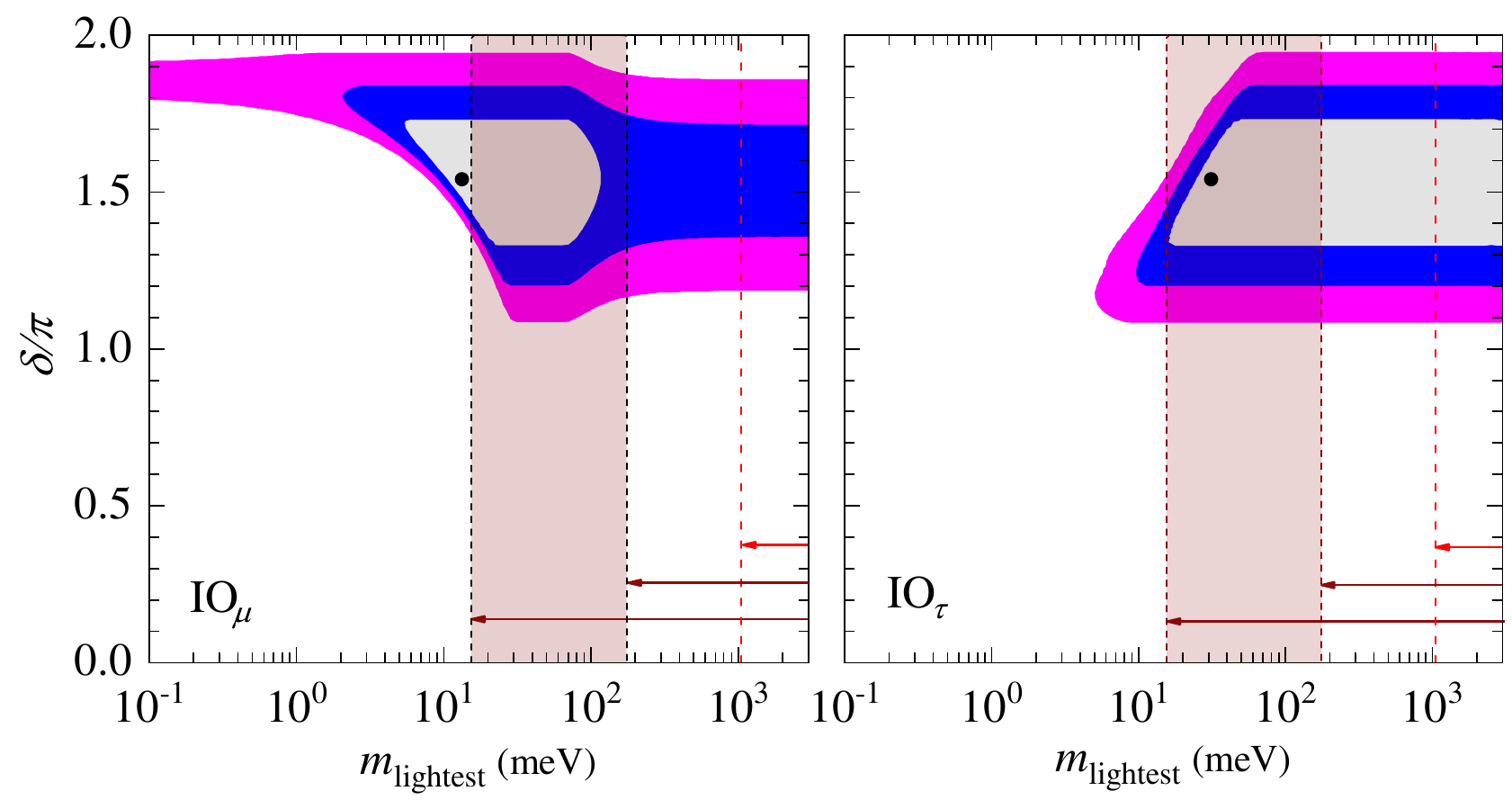}
\caption{Allowed regions in ($m_\text{lightest}$,$\delta$) for NO (upper plots) and IO (lower plots) for the $\mathcal{Z}_8$ model. The colour code for the allowed regions is the same as in fig.~\ref{results_t23}. The vertical shaded region corresponds to the upper-limit range for $m_\text{lightest}$ coming from the cosmological bound on $\sum_k m_k$ (see text). The vertical dashed red line shows the $m_\text{lightest}$ upper limit from the $m_{\beta}<1.1$~eV KATRIN limit (90\% CL).}
\label{results_ml}
\end{figure}

In figs.~\ref{results_t23} and~\ref{results_ml} we present the $1\sigma$, $2\sigma$ and $3\sigma$ allowed regions in the ($\theta_{23}$,$\delta$) and ($m_{\rm lightest}$,$\delta$) planes, for decoupled $e$, $\mu$ and $\tau$ schemes, with both NO (upper panels) and IO (lower panels) neutrino mass spectra. These distinct cases are labeled as NO$_{e,\mu,\tau}$ and IO$_{e,\mu,\tau}$, respectively. To compute the $\chi^2$ we use the one-dimensional global-fit profiles also from~\cite{deSalas:2020pgw} for $s_{12}^2$, $s_{13}^2$, $\Delta m_{21}^2$ and $\Delta m_{31}^2$. Due to the correlations observed between $\theta_{23}$ and $\delta$, we use instead for these two parameters a two-dimensional $\chi^2$ distribution, also from Ref.~\cite{deSalas:2020pgw}. Notice that we do not include in the fit the constraints on $m_{\rm lightest}$ coming from $\beta$-decay experiments or cosmology. Instead, in fig.~\ref{results_ml} we display our results in terms of $m_{\rm lightest}$ and indicate these bounds by vertical red dashed line and a shaded band, respectively. The vertical dashed red line corresponds to the $m_\text{lightest}$ KATRIN tritium beta decay upper limit $m_{\beta}<1.1$~eV (90\% CL). On the other hand the band delimits the conservative and agressive upper limits from cosmology, as indicated in the caption. The right (left) delimiting black dotted limit corresponds to the Planck TT+lowE (Planck TT, TE, EE+lowE+lensing+BAO) 95\% CL limit $\sum_k m_k<0.54$~eV ($0.12$~eV). Notice that the fitting procedure is always performed under the theoretical assumption expressed by relation \eqref{c11} for each of the three cases to be analysed. We also include in fig.~\ref{results_t23} the (dashed) lines delimiting the regions implied by data only, i.e. without any ``prior'' input from the model.
\begin{figure}[t!]
\includegraphics[width=1.0\textwidth]{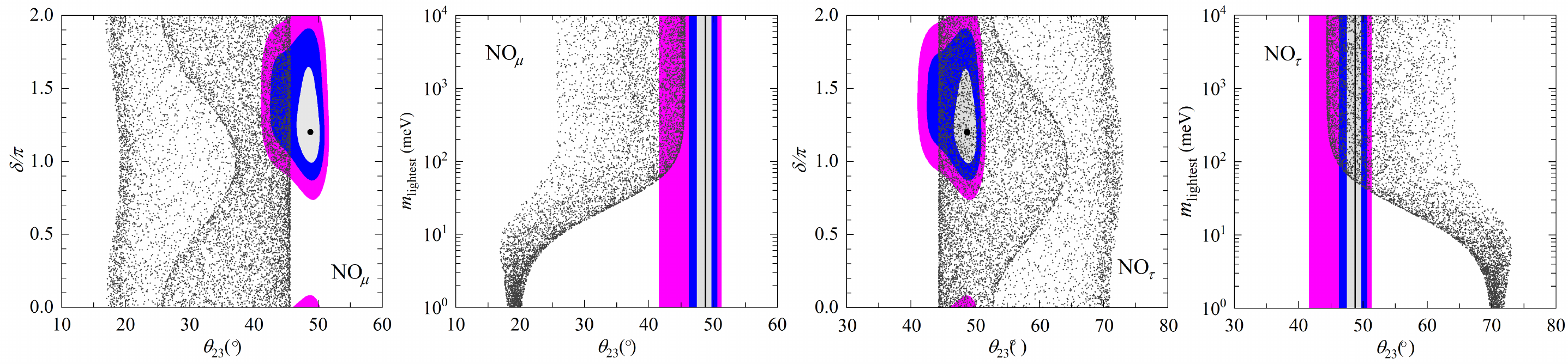}
\caption{Model preference for the first (second) $\theta_{23}$ octant in the NO$_\mu$ (NO$_\tau$) scenario. The $(\theta_{23},\delta)$ scatter points (first and third panels) and $(\theta_{23},m_\text{lightest})$ (second and fourth panels) obey eq.~\eqref{c11}. The $1\sigma$, $2\sigma$ and $3\sigma$ regions are shown in gray, blue and magenta, respectively. The black dot and vertical line correspond to the best-fit.  The overlap between the scatter plots and the experimentally-allowed regions produce the results of fig.~\ref{results_t23}.}
\label{NOmuNOtau}
\end{figure}

Examining these figures we conclude the following:
\begin{itemize}

\item The IO$_e$ case is not compatible with data since the condition $(\mathbf{M}_\nu')_{11}=0$ leads to a vanishing \znbb decay rate. The latter is well-known to be inconsistent with inverse ordering.

\item As seen in the leftmost upper panel in fig.~\ref{results_t23}, for the NO$_e$ case the model-allowed regions in ($\theta_{23}$,$\delta$) coincide with the generic ones obtained in the global fit of experimental data. This is due to the fact that $(\mathbf{M}_\nu')_{11}$ does not depend on $\theta_{23}$ and $\delta$. However, $(\mathbf{M}_\nu')_{11}=0$ is only satisfied in the $(m_{\rm lightest},\delta)$ allowed regions shown in the corresponding plot of fig.~\ref{results_ml}. In particular, compatibility with data in the NO$_e$ case requires $m_{\rm lightest} \sim [20,80]$~meV, at the $3\sigma$ level. This leads to a lower bound on $m_{\rm lightest}$. Moreover, it implies an upper bound on $m_{\rm lightest}$ which is more stringent than those which follow from current cosmology and the kinematics of tritium beta decays. These are indicated in fig.~\ref{results_ml} by a vertical shaded band and a red dashed line, respectively.
\item The NO$_\mu$ (NO$_\tau$) case selects the first (second) octant for $\theta_{23}$, as can be seen in the middle (rightmost) upper plot of fig.~\ref{results_t23}. This can also be seen in fig.~\ref{NOmuNOtau}. Consequently, while NO$_\tau$ is compatible with data at $1\sigma$, for NO$_\mu$ this happens only at $2\sigma$. The preference for a different $\theta_{23}$ octant in each of these two cases is interesting since future experiments, such as T2HK, might in principle resolve the ambiguity and, hence exclude either NO$_\mu$ or NO$_\tau$.
\item For the IO$_{\mu,\tau}$ cases the $(\theta_{2,3},\delta)$ regions coincide with those given by the current global fits, see lower panels in figs.~\ref{results_t23}, indicating that the model can accommodate neutrino oscillation data within the corresponding $(m_{\rm lightest},\delta)$ regions (lower panels in fig.~\ref{results_ml}). One sees that for IO$_{\mu}$ the $m_{\rm lightest}$ range is $[5,100]$~meV at $1\sigma$, with the best-fit value at $\sim 10$~meV. As for IO$_{\tau}$, there is a $1\sigma$ lower bound of $16$~meV, which lies in the left border of the cosmological bound band.
\end{itemize}    

\section{neutrinoless double beta decay predictions }
\label{sec:dbd}
\begin{figure}[t!]
	\includegraphics[width=0.7\textwidth]{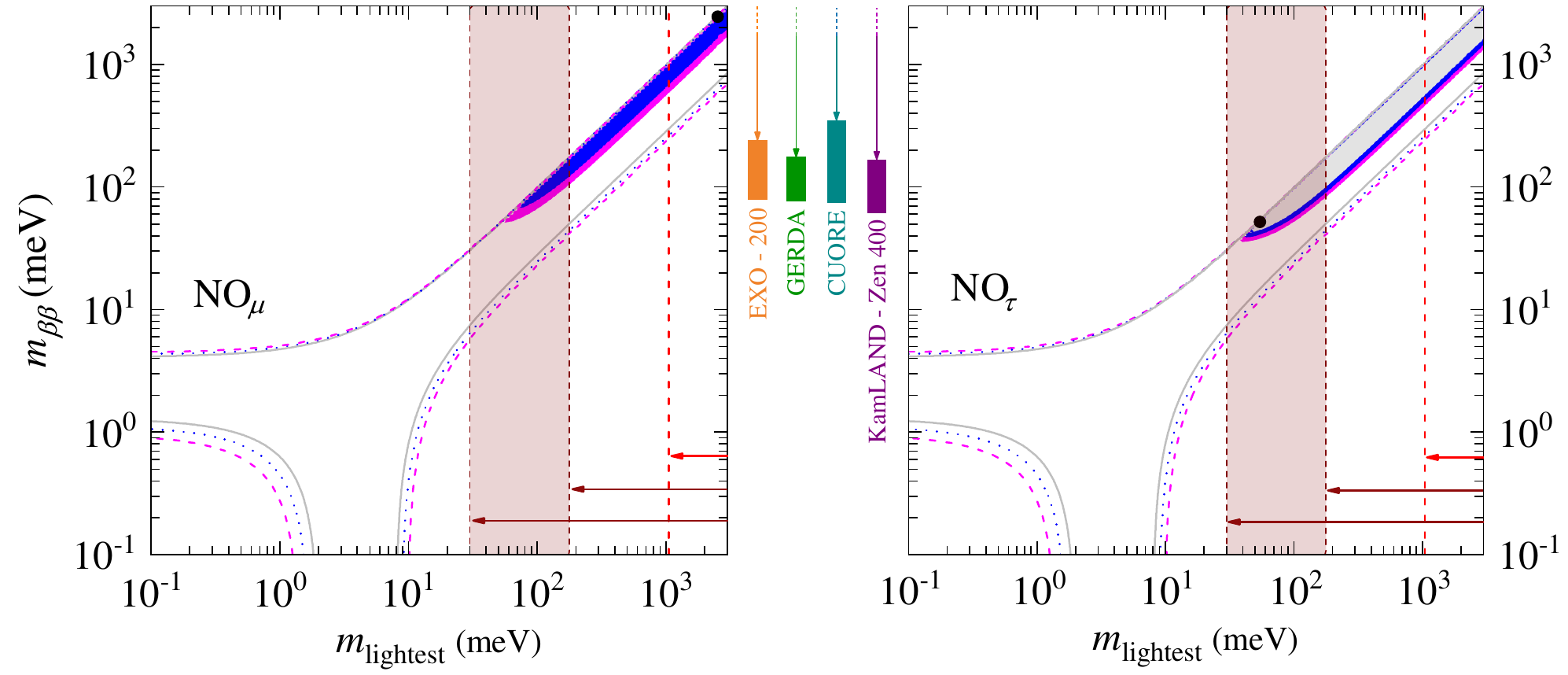}\\
	\includegraphics[width=0.7\textwidth]{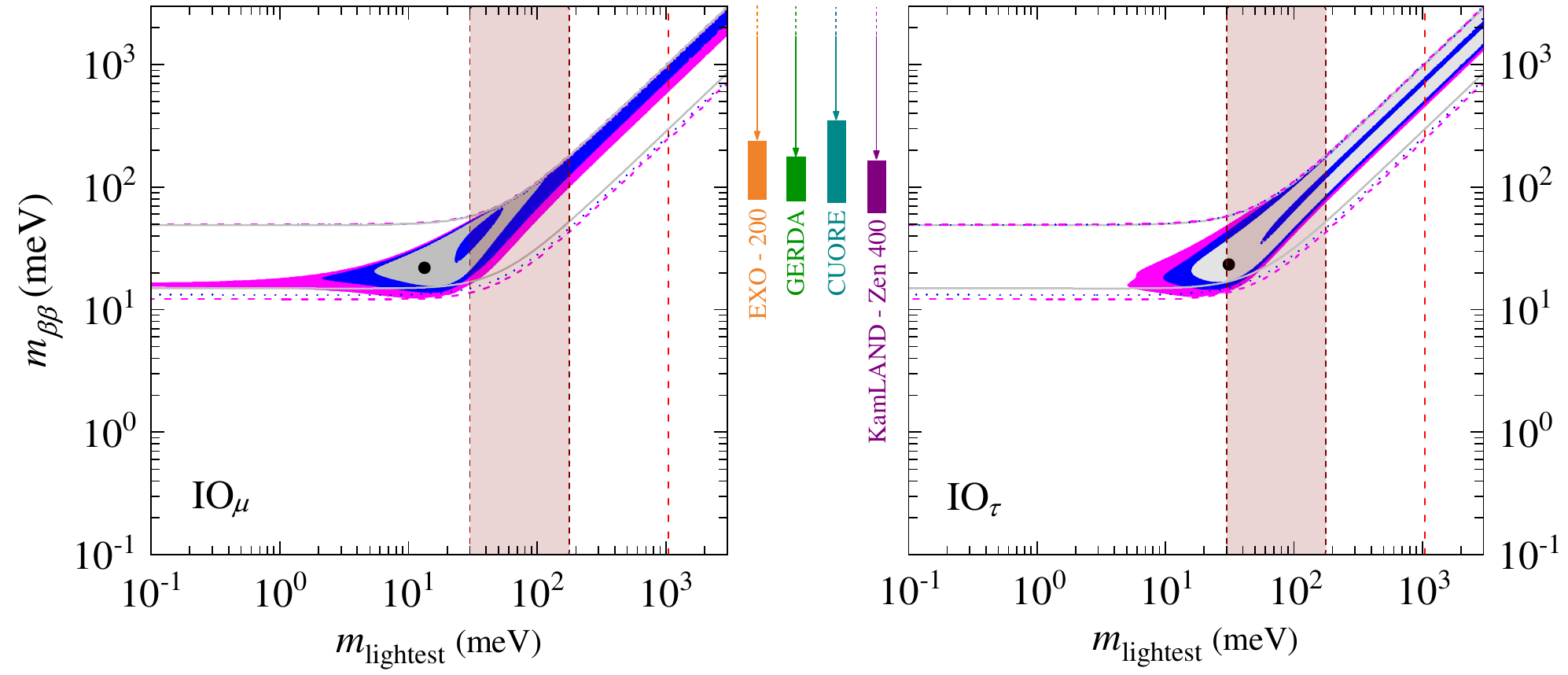}
	\caption{Allowed ($m_\text{lightest}$,$m_{\beta\beta}$) regions for NO (upper panels) and IO (lower panels). The colour code is the same in figs.~\ref{results_t23} and \ref{results_ml}. The solid (dotted) [dashed] lines delimit the $1\sigma$ ($2\sigma$) [$3\sigma$] $m_{\beta\beta}$ regions allowed in the general unconstrained case. The vertical bars in between the panels indicate the current $m_{\beta\beta}$ upper bounds from KamLAND-Zen~400~\cite{KamLAND-Zen:2016pfg},
          GERDA~\cite{Agostini:2020xta}, CUORE~\cite{Adams:2019jhp} and EXO-200~\cite{Anton:2019wmi} at 95\%CL. The height of the bars reflects the uncertainties in the nuclear matrix elements relevant for the computation of the decay rates.}
	\label{results_mbb}
\end{figure}

Given the results just obtained for the oscillations parameters and the lightest neutrino mass, we now present the allowed regions for the effective mass parameter $m_{\beta\beta}$ characterizing the amplitude for neutrinoless double beta decay. In the symmetrical parametrization of the lepton mixing matrix~\cite{Schechter:1980gr,Rodejohann:2011vc} the contribution arising from the exchange of light neutrinos gives 
\begin{align}
{\rm NO}: \;&m_{\beta\beta}=\left| c_{12}^2 c_{13}^2  \,m_{\rm lightest} + s_{12}^2 c_{13}^2  \,\sqrt{m_{\rm lightest}^2+\dmsol} \,\text{e}^{2 i \phi_{12}}+
s_{13}^2  \, \sqrt{m_{\rm lightest}^2+\dmatm}  \,\text{e}^{2 i \phi_{13}}
\right| \,,\\
{\rm IO}: \;&m_{\beta\beta}=\left| c_{12}^2 c_{13}^2  \,\sqrt{m_{\rm lightest}^2+\dmsol}  + s_{12}^2 c_{13}^2  \,\sqrt{m_{\rm lightest}^2+\dmsol+|\dmatm|}  \,\text{e}^{2 i \phi_{12}}+
s_{13}^2  \, m_{\rm lightest}  \,\text{e}^{2 i \phi_{13}}
\right| \,,
\end{align}
for the NO and IO cases respectively.

In fig.~\ref{results_mbb} we show the predictions for $m_{\beta\beta}$ taking into account the global oscillation analysis in Ref.~\cite{deSalas:2020pgw}. The results are given in terms of $m_{\rm lightest}$ and correspond to those of figs.~\ref{results_t23} and \ref{results_ml}. The upper panels correspond to the NO case, while the lower ones are for the IO. The present upper bounds coming from \znbb searches at KamLAND-Zen 400~\cite{KamLAND-Zen:2016pfg}, GERDA~\cite{Agostini:2020xta}, CUORE~\cite{Adams:2019jhp} and
EXO-200~\cite{Anton:2019wmi} are also indicated. Note that the case NO$_{e}$ predicts $m_{\beta\beta}=0$ and is not shown. From the results presented in that figure one sees that there is a lower bound on the \znbb amplitude that holds even if the neutrino mass spectrum is normal ordered. Although this feature occurs in many family symmetry schemes where the cancellation is prevented by the structure of the leptonic weak interaction vertex~\cite{Dorame:2011eb,Dorame:2012zv,King:2013hj}, it is remarkable that it holds here, with just such a simple Abelian $\mathcal{Z}_8$ symmetry. In fact, one sees that the current KamLAND bound for $m_{\beta\beta}$ nearly excludes the NO cases. Future data from upcoming projects like AMORE II~\cite{Lee:2020rjh}, CUPID~\cite{Wang:2015raa}, LEGEND~\cite{Abgrall:2017syy}, SNO+~I~\cite{Andringa:2015tza},
KamLAND2-Zen~\cite{KamLAND-Zen:2016pfg}, nEXO~\cite{Albert:2017hjq} or PandaX-III~\cite{Chen:2016qcd} should probe the entire allowed regions. For the inverted ordering cases one sees that the allowed regions are also meaningfully probed by the current $m_{\beta\beta}$ upper bound. However, they will be fully probed by the expected sensitivities from upcoming experiments.

\section{Conclusions}
\label{sec:concl}

We have proposed a simple scoto-seesaw model based on a $\mathcal{Z}_8$ flavour symmetry which is broken to a  {\em dark} $\mathcal{Z}_2$ by a scalar singlet $\sigma$ VEV. This complex vacuum expectation value is the unique source of leptonic CP violation, that takes place in a spontaneous manner. Such CPV is communicated to the leptonic sector via couplings of $\sigma$ with the $\nu_{R}$ and the dark fermion $f$. We have shown that the $\mathcal{Z}_8$ symmetry leads to constraints on the low-energy parameters. Except for the case in which $m_{\beta\beta}=0$, the predicted ranges on the lightest-neutrino mass for a normal ordered neutrino spectrum will be fully tested by future neutrinoless double beta decay experiments and by improved neutrino mass sensitivities from cosmological probes. In the inverted ordering case, a better determination of the Dirac CPV phase $\delta$ is required to test the model under scrutiny. More than proposing yet another model for neutrino masses and dark matter, our work intends to establish a template for a dynamical origin of leptonic CP violation. This is achieved by combining the neutrino mass generation mechanism and the solution of the dark matter problem. CP violation is driven by the same scalar singlet VEV providing mass to the heavy light-neutrino mass mediator fermions.

\begin{acknowledgments}
We thank G.C. Branco and M. Tortola for discussions. This work is supported by the Spanish grants SEV-2014-0398 and FPA2017-85216-P (AEI/FEDER, UE), PROMETEO/2018/165 (Generalitat Valenciana), the Spanish Red Consolider MultiDark FPA2017-90566-REDC and Funda\c{c}{\~a}o para a Ci{\^e}ncia e a Tecnologia (FCT, Portugal) through the projects UIDB/00777/2020, UIDP/00777/2020, CERN/FIS-PAR/0004/2019, and PTDC/FIS-PAR/29436/2017. D.M.B. is supported by the FCT grant SFRH/BD/137127/2018. R.S. is supported by SERB, Government of India grant SRG/2020/002303.

\end{acknowledgments}
\newpage
\appendix*
\section{The scalar sector of the $\mathcal{Z}_8$ model, SCPV and scotogenic loop(s)}
\label{SCPV}

In this work we explore the possibility of having SCPV in the $\mathcal{Z}_8$ scoto-seesaw model characterised by the field content and transformation properties shown in table~\ref{modelZ8}. Imposing CP invariance at the Lagrangian level, the most general invariant scalar potential is
\begin{align}
	V=\;&m_\Phi^2\Phi^\dagger \Phi+m_\eta^2\eta^\dagger\eta+m_\sigma^2\sigma^\ast \sigma+\dfrac{m'^2_\sigma}{2}(\sigma^2+\sigma^{\ast 2})+m_\chi^2\chi^\ast \chi\nonumber\\
	&+\mu_1(\chi^2\sigma+\chi^{\ast 2}\sigma^{\ast})+\mu_2(\eta^\dagger \Phi \chi^\ast+\Phi^\dagger \eta \chi)\nonumber\\
	&+\dfrac{\lambda_1}{2}(\Phi^\dagger \Phi)^2+\dfrac{\lambda_2}{2}(\eta^\dagger \eta)^2+\dfrac{\lambda_{3}}{2}(\sigma^\ast \sigma)^2+\dfrac{\lambda_4}{2}(\chi^\ast \chi)^2+\dfrac{\lambda'_{3}}{4}(\sigma^4+\sigma^{\ast4})+\lambda_{5}(\Phi^\dagger \Phi)(\eta^\dagger \eta)+\lambda'_{5}(\Phi^\dagger \eta)(\eta^\dagger \Phi)\nonumber\\
	&+\lambda_{6}(\Phi^\dagger\Phi)(\sigma^\ast\sigma)+\lambda_{7}(\Phi^\dagger\Phi)(\chi^\ast\chi)+\lambda_{8}(\eta^\dagger\eta)(\sigma^\ast\sigma)+\lambda_{9}(\eta^\dagger\eta)(\chi^\ast\chi)+\lambda_{10}(\sigma^\ast\sigma)(\chi^\ast\chi)\nonumber\\
	&+\lambda_{11}(\eta^\dagger\Phi\sigma\chi+\Phi^\dagger\eta\sigma^\ast\chi^\ast)\,,
\end{align}
where all parameters are real. Notice that the $m'^2_\sigma$ breaks the $\mathcal{Z}_8$ symmetry softly, avoiding in this way the formation of cosmological domain walls due to spontaneous breaking of a discrete symmetry~\footnote{For alternative solutions to this problem in the context of neutrino mass models see Ref.~\cite{Lazarides:2018aev}}. We parameterise the scalar fields of the model as
\begin{align}
	\Phi=\begin{pmatrix}
		\phi^+\\
		\dfrac{v+\phi_{0\text{R}}+i\phi_{0\text{I}}}{\sqrt{2}}
	\end{pmatrix},\;	
	\eta=\begin{pmatrix}
		\eta^+\\
		\dfrac{v_\eta e^{i\theta_\eta}+\eta_{0\text{R}}+i\eta_{0\text{I}}}{\sqrt{2}}
	\end{pmatrix},\;
	\chi=\dfrac{v_\chi +\chi_{\text{R}}+i\chi_{\text{I}}}{\sqrt{2}},\;
	\sigma=\dfrac{u e^{i\theta} +\sigma_{\text{R}}+i\sigma_{\text{I}}}{\sqrt{2}}\,,
\end{align}
with the corresponding vacuum configurations given by
\begin{align}
	\langle\Phi\rangle=\begin{pmatrix}
		0\\
		\dfrac{v}{\sqrt{2}}
	\end{pmatrix},\;	
	\langle\eta\rangle=\begin{pmatrix}
		0\\
		\dfrac{v_\eta e^{i\theta_\eta}}{\sqrt{2}}
	\end{pmatrix},\;
	\langle\chi\rangle=\dfrac{v_\chi }{\sqrt{2}},\;
	\langle\sigma\rangle=\dfrac{u e^{i\theta}}{\sqrt{2}}\,,
\end{align}
As explained in Section~\ref{sec:scot-dm-neutr}, we are interested in potential minima with vanishing VEVs for the \emph{dark} scalars, i.e. $v_\eta=v_\chi=0$ and $u,v\neq 0$. In such cases, the three nontrivial minimization conditions are
\begin{align}
	m_\Phi^2&+\dfrac{\lambda_1}{2}v^2+\dfrac{\lambda_{6}}{2}u^2=0\;,\label{dVdvphi1}\\
	m_\sigma^2&+m'^2_\sigma \cos (2\theta) +u^2\left[\dfrac{\lambda_3}{2}+\dfrac{\lambda'_3}{2}\cos (4\theta)\right]+\dfrac{\lambda_{6}}{2}v^2=0\;,\label{dVdvsigma1}\\
	m'^2_\sigma& \sin (2\theta)+\dfrac{\lambda'_3}{2}u^2\sin (4\theta)=0\;,\label{dVdtsigma1}	
\end{align}
for $v,u\neq 0$. The above equations can be solved for $m_\Phi^2$, $m_\sigma^2$ and $\theta$ as function of $u,v$ and the remaining potential parameters. Namely,
\begin{align}
	(\text{S}1)&\quad m_\Phi^2=-\dfrac{\lambda_1}{2}v^2-\dfrac{\lambda_6}{2}u^2\;, \quad m_\sigma^2=-\dfrac{\lambda_6}{2}v^2-\dfrac{\lambda_3-\lambda_3'}{2}u^2\;,\quad\cos (2\theta)=-\dfrac{m'^2_\sigma}{u^2\lambda'_3}\;;\label{potminimum}\\
	(\text{S}2)&\quad m_\Phi^2=-\dfrac{\lambda_1}{2}v^2-\dfrac{\lambda_6}{2}u^2\;, \quad m_\sigma^2=-m_\sigma'^2-\dfrac{\lambda_6}{2}v^2-\dfrac{\lambda_3+\lambda_3'}{2}u^2\;,\quad\theta=k \pi,\;k\in\mathcal{Z}\;; \\
	(\text{S}3)&\quad m_\Phi^2=-\dfrac{\lambda_1}{2}v^2-\dfrac{\lambda_6}{2}u^2\;, \quad m_\sigma^2=m_\sigma'^2-\dfrac{\lambda_6}{2}v^2-\dfrac{\lambda_3+\lambda_3'}{2}u^2\;,\quad\theta=\dfrac{\pi}{2}+ k\pi,\;k\in\mathcal{Z}\;.
\end{align}
It can be shown that only case (S1) leads to spontaneous CP violation since for (S2) and (S3) a CP transformation can be defined such that there is full CP invariance. Notice that if $m_\sigma'^2=0$ (which corresponds to having an exact $\mathcal{Z}_8$ symmetric scalar potential), $\theta=\pi/4$ which is still a CPV solution. However in this case the $\mathcal{Z}_8$ is spontaneously broken leading to the formation of cosmological domain walls. In order for (S1) to correspond to the deepest minimum among the above three, the condition $(m_\sigma'^4-u^4\lambda_3'^2)/(4\lambda_3')>0$ must be verified.

The masses of the \emph{dark} charged scalars are $m^2_{\eta^{\pm}}=m_\eta^2+\lambda_5v^2/2+\lambda_8u^2/2$, while the \emph{non-dark} and \emph{dark} neutral scalar mass matrices $\mathcal{M}^2_{\phi\sigma}$ and $\mathcal{M}^2_{\eta\chi}$, respectively, in the $(\phi_{0\text{R}},\sigma_\text{R},\sigma_\text{I})$ and $(\eta_{0\text{R}},\chi_\text{R},\eta_{0\text{I}},\chi_\text{I})$ basis read
\begin{align}
	\mathcal{M}^2_{\phi\sigma}=\begin{pmatrix}
		v^2\lambda_1&v u\lambda_6\cos\theta&v u\lambda_6\sin\theta\\
		\cdot&u^2(\lambda^3+\lambda'_3)\cos^2\theta&u^2(\lambda_3-3\lambda'_3)\cos\theta\sin\theta\\
		\cdot&\cdot&u^2(\lambda^3+\lambda'_3)\sin^2\theta\\
	\end{pmatrix},
\end{align}
\begin{figure}[t!]
	\includegraphics[width=0.3\textwidth]{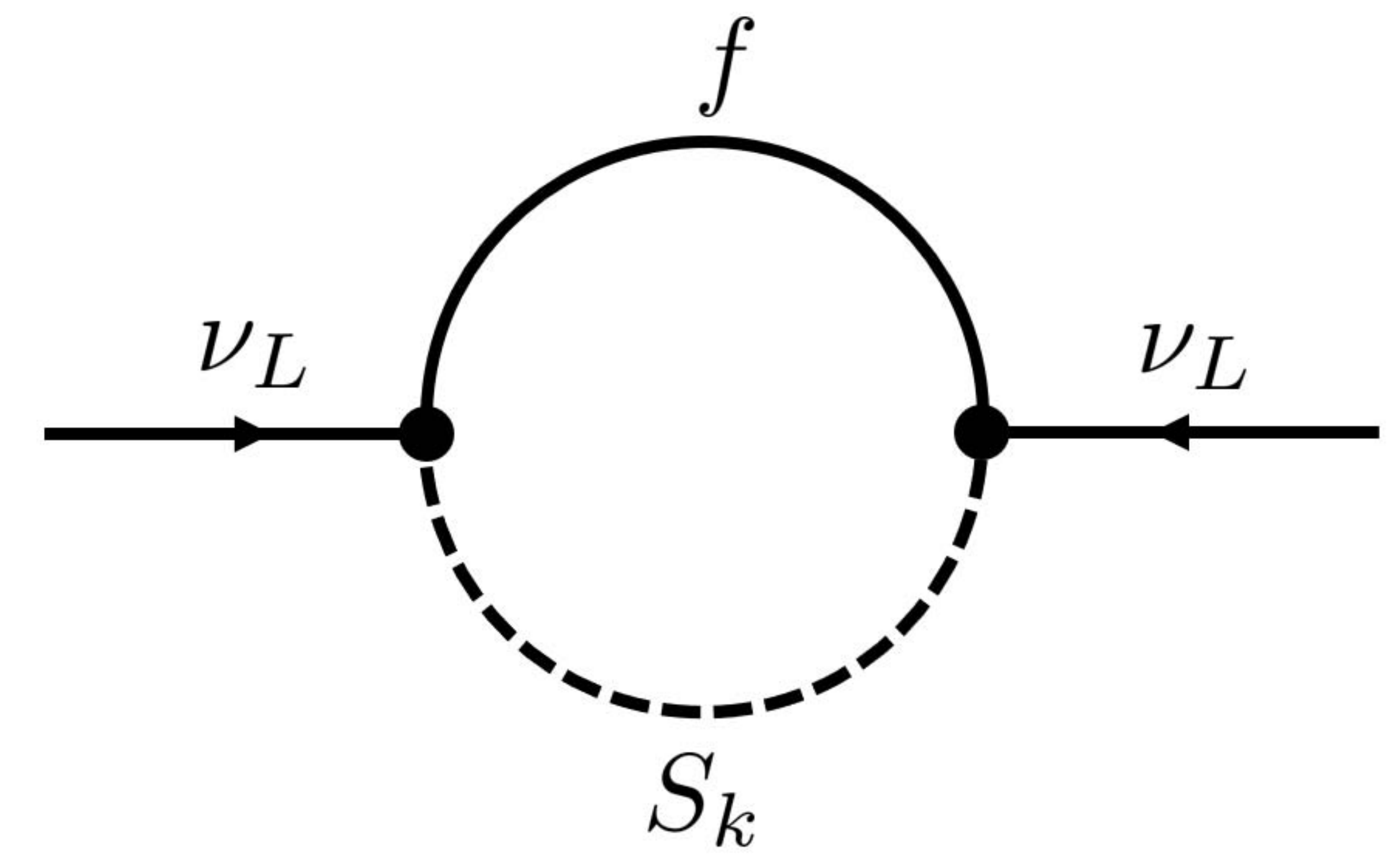}
	\caption{Scotogenic loop with scalar mass eigenstate basis, equivalent to the ones of fig.~\ref{scotoloopsZ8} given in the gauge eigenfield basis.}
	\label{figloop}
\end{figure}
\scalebox{0.90}{\parbox{\linewidth}{%
		\begin{align}
			\mathcal{M}^2_{\eta\chi}=\begin{pmatrix}
				m_\eta^2+\frac{\lambda_5+\lambda'_5}{2}v^2+\frac{\lambda_8}{2}u^2&v\left(\frac{\mu_2}{\sqrt{2}}+\frac{\lambda_{11}}{2}u\cos\theta\right)&0&-\frac{\lambda_{11}}{2}v u \sin\theta\\
				\cdot&m_\chi^2+\frac{\lambda_7}{2}v^2+\frac{\lambda_{10}}{2}u^2+\sqrt{2}u\lambda_{11}\cos\theta&\frac{\lambda_{11}}{2}v u \sin\theta&-\sqrt{2}\mu_1 u\sin\theta\\
				\cdot&\cdot&m_\eta^2+\frac{\lambda_5+\lambda'_5}{2}v^2+\frac{\lambda_8}{2}u^2&v\left(-\frac{\mu_2}{\sqrt{2}}+\frac{\lambda_{11}}{2}u\cos\theta\right)\\
				\cdot&\cdot&\cdot&m_\chi^2+\frac{\lambda_7}{2}v^2+\frac{\lambda_{10}}{2}u^2-\sqrt{2}u\lambda_{11}\cos\theta\\
			\end{pmatrix}.
			\label{etachimix}
		\end{align}
}}

The \emph{dark} scalar mass eigenstates $S_i$ with masses $m_{S_i}$ ($i=1,2,3,4$) are the ones relevant for the one-loop scotogenic neutrino mass matrix. The neutral components of $\eta$ and $\chi$ are related to $S_i$ through a unitary matrix $\mathbf{V}$ such that
\begin{align}
   \begin{pmatrix}
    \eta_{0\text{R}}\\
    \chi_\text{R}\\
    \eta_{0\text{I}}\\
    \chi_\text{I}
    \end{pmatrix}=\mathbf{V}\begin{pmatrix}
    S_1\\
    S_2\\
    S_3\\
    S_4
    \end{pmatrix}\;,\; \eta=\sum_{k=1}^4(\mathbf{V}_{1k}+i \mathbf{V}_{3k})\,S_k\,.
\end{align}
The scotogenic loop factor in the mass-eigenstate basis of the \emph{dark} fermion $f$ and scalars $S_k$ corresponding to the ones in the gauge basis (see fig.~\ref{scotoloopsZ8}) is shown in fig.~\ref{figloop}. The loop factor relevant for the scotogenic neutrino mass is
\begin{align}
	\mathcal{F}(M_f,S_k)=\dfrac{1}{32\pi^2}\sum_{k=1}^4(\mathbf{V}_{1k}-i \mathbf{V}_{3k})^2\dfrac{\,m_{S_k}^2}{M_f^2-m_{S_k}^2}\log\left(\dfrac{M_f^2}{m_{S_k}^2}\right).
\end{align}
As it should, $\mathcal{F}(M_f,S_k)=0$ in the limit $\mu_{1,2}=0$ and $\lambda_{11}=0$ since in this case the scalar and pseudoscalar components of $\eta$ and $\chi$ are degenerate among themselves and do not mix (this can be readily seen fig.~\ref{scotoloopsZ8}).

\bibliographystyle{utphys}
\bibliography{bibliography}

\end{document}